\documentclass[floatfix, preprint, onecolumn, prd,showpacs, showkeys, preprintnumbers, nofootinbib, superscriptaddress]{revtex4-1}
\usepackage{times}
\usepackage{amssymb,amsbsy,amsmath,amsfonts}
\usepackage{graphicx}
\usepackage{float}
\usepackage{color}
\usepackage{morefloats}
\usepackage{rotating}
\usepackage{srcltx}
\usepackage{slashed}
\usepackage{verbatim}
\usepackage{tabularx}

\begin{document}

\title{Pseudocalar meson and vector meson interactions and dynamically generated axial-vector mesons}

\author{Yu Zhou}
\affiliation{School of Physics and Nuclear Energy Engineering and International Research Center for Nuclei and Particles
in the Cosmos, Beihang University, Beijing 100191, China}

\author{Xiu-Lei Ren}
\affiliation{School of Physics and Nuclear Energy Engineering and International Research Center for Nuclei and Particles
in the Cosmos, Beihang University, Beijing 100191, China}

\author{Hua-Xing Chen}
\affiliation{School of Physics and Nuclear Energy Engineering and International Research Center for Nuclei and Particles
in the Cosmos, Beihang University, Beijing 100191, China}

\author{Li-Sheng Geng}
\email[E-mail me at: ]{lisheng.geng@buaa.edu.cn}
\affiliation{School of Physics and Nuclear Energy Engineering and International Research Center for Nuclei and Particles
in the Cosmos, Beihang University, Beijing 100191, China}

\begin{abstract}
The axial-vector mesons $a_1(1260)$, $b_1(1235)$, $f_1(1285)$, $h_1(1170)$, $h_1(1380)$,  and $K_1(1270)$ 
are dynamically generated in the unitized chiral perturbation theory. Such a picture has been
tested extensively  in the past few years. In this work, we calculate
the interaction kernel up to $\mathcal{O}(p^2)$ and study the impact on the dynamically generated axial-vector states. In anticipation of future lattice QCD simulations,
we calculate the scattering lengths and the pole positions as  functions of the pion mass, with the light-quark mass dependence of the kaon, the eta, and the vector mesons 
determined by the $n_f=2+1$ lattice QCD simulations of the PACS-CS Collaboration.
\end{abstract}


\date{\today}

\maketitle
\section{Introduction}
Understanding the quark contents of hadrons has been at the forefront of nonperturbative strong interaction physics. In the naive quark model,
baryons are composed of three quarks or antiquarks, and mesons consist of a pair of quark and antiquark. This picture works extremely well
for most hadrons discovered before 2000, except a few, such as the scalar nonet of the $f_0(500)$~\cite{Jaffe:1976ig} and the $\Lambda(1405)$~\cite{Dalitz:1959dn}. On the other hand,
in the past decade many newly discovered resonances cannot be easily accommodated by the naive quark model. Some of them clearly
contain more than the minimum number of valence quarks, such as the $Z_c(3900)$~\cite{Ablikim:2013mio,Liu:2013dau}, while others may have components of both $q\bar{q}$ ($qqq$) and multi quark configurations. Although a vast amount of experimental and theoretical studies have been performed to understand  their nature, much more remain to be done.

Even among the seemingly well established and understood hadrons, some of them may be more complicated than originally thought. One such example is the lowest-lying axial-vector mesons. It was shown in Refs.~\cite{Lutz:2003fm,Roca:2005nm} that the $a_1(1260)$, $b_1(1235)$, $f_1(1285)$, $h_1(1170)$, $h_1(1380)$, and $K_1(1270)$ can be built from the interactions between
the pseudoscalar octet of the pion and the vector nonet of the $\rho$ within the so-called unitized chiral perturbation theory (UChPT) approach (see Refs.~\cite{  
  Kaiser:1995eg,Dobado:1996ps,Oller:1997ti,Oset:1997it,Oset:1997it,Oller:1998hw,Kaiser:1998fi,Oller:1998zr,Oller:2000fj,Lutz:2001yb} for some early references). There are some technical differences between the two approaches. In Ref.~\cite{Lutz:2003fm}, both the $K_1(1270)$ and the $K_1(1400)$ are claimed to be dynamically generated while in Ref.~\cite{Roca:2005nm} the two poles found in the isospin 1/2 and strangeness 1 channel are both claimed to be the $K_1(1270)$--the so-called two-pole picture.\footnote{The two-pole scenario is also found for the $\Lambda(1405)$~\cite{Oller:2000fj,Jido:2003cb}, which seems to be a universal feature of all the studies based on coupled channel chiral dynamics. See Ref.~\cite{Hyodo:2011ur} for a recent review.} Such a dynamical picture has been put into test in many different scenarios, e.g., the two-pole structure of the $K_1(1270)$~\cite{Geng:2006yb}, their radiative decays~\cite{ Roca:2006am,Lutz:2008km,Nagahiro:2008zza,Nagahiro:2008cv},  their large $N_c$ behavior~\cite{Geng:2008ag},  the composite and elementary nature of the $a_1(1260)$~\cite{Nagahiro:2011jn} and its finite volume dependence~\cite{,Roca:2012rx}. All these studies suggest that the axial-vector mesons contain large pseudoscalar meson-vector meson components.

It should be noted that the studies of Refs.~\cite{Lutz:2003fm,Roca:2005nm} are both based on the leading-order (LO) chiral potentials. It is not clear whether the scenario of these states being dynamically generated will change when higher order kernels are included. Given the fact that lately there have been attempts to study these states on the lattice~\cite{Prelovsek:2011im,Lang:2014tia}, such a study is urgently needed. Recently it has been shown  in the $DK$ sector that although inclusion of higher-order kernels do not change qualitatively the conclusions, they do have impact on the quantitative description of the lattice chromodynamics (LQCD) data, more specifically, their dependence on the light-quark masses (see, e.g., Ref.~\cite{Altenbuchinger:2013vwa}). The main purpose of the present work is to include the contribution of the next-to-leading order (NLO) chiral potentials, and to study their impact on the LO results and on the light-quark mass dependence of the scattering lengths and the pole positions.

The paper is organized as follows. In Sec. 2, we briefly recall the basic framework of UChPT in studies of the interactions between the pseudoscalar octet of the pion and the vector nonet of the $\rho$, and introduce the NLO chiral potentials. We study their impact on the prediction of the dynamically generated axial-vector mesons in Sec.3.  In anticipation of future studies of these resonances on the lattice, we calculate the scattering lengths of the pseudoscalar-vector meson interactions and study their light-quark mass dependence in Sec. 4.  The pole positions and their light-quark mass dependence are studied in Sec. 5. A short summary is given in Sec.6.

\section{Theoretical Framework}
UChPT has become an invaluable tool to study interactions between hadrons and has played an important role in helping understand
the nature of the many newly discovered resonances. At the center of the UChPT is the kernel provided by chiral dynamics and a unitarization procedure.
Although chiral dynamics largely fix the form of the kernel, unitarization techniques differ. Nevertheless, various unitarization techniques generally lead to similar results and only fine details may differ.  In both Refs.~\cite{Lutz:2003fm,Roca:2005nm}, the  Bethe-Salpeter equation approach was adopted to unitarize the chiral kernel, which has the following schematic form:
\begin{equation}
T=V+VGT,
\end{equation}
where $V$ is the kernel potential, $T$ the scattering $T$-matrix, and $G$ the one-loop scalar two-point function.

\subsection{LO chiral potentials}
The LO amplitudes of pseudoscalar meson-vector meson interactions are calculated with the following interaction Lagrangian~\cite{Lutz:2003fm,Roca:2005nm}:
\begin{equation}
\label{Lagrangian}
\mathcal{L}_I =-\frac{1}{4}\textrm{Tr}\{(\nabla_\mu V_\nu-\nabla_\nu V_\mu)(\nabla^\mu V^\nu-\nabla^\nu V^\mu)\},
\end{equation}
where Tr means SU(3) trace and $\nabla_\mu$ is the covariant derivative defined as
\begin{equation}
\nabla_\mu V_\nu=\partial_\mu V_\nu+[\Gamma_\mu,V_\nu],
\end{equation}
where $[,]$ stands for commutator and $\Gamma_\mu$ is the vector current
\begin{equation}
\Gamma_\mu=\frac{1}{2}(u^\dag\partial_\mu u+u\partial_\mu u^\dag)
\end{equation}
with
\begin{equation}
u^2=U=e^{i\frac{\sqrt{2}}{f}P}.
\end{equation}
In the previous equations $f$ is the pion decay constant in the chiral limit and $P$ and $V$ are the SU(3) matrices containing the octet of pseudoscalar and the nonet of vector mesons, respectively,
\begin{equation}
P=\left(
    \begin{array}{ccc}
      \frac{1}{\sqrt{2}}\pi^0+\frac{1}{\sqrt{6}}\eta_8 & \pi^+ & K^+ \\
      \pi^- & -\frac{1}{\sqrt{2}}\pi^0+\frac{1}{\sqrt{6}}\eta_8 & K^0 \\
      K^- & \bar{K}^0 & -\frac{2}{\sqrt{6}}\eta_8 \\
    \end{array}
  \right),
\end{equation}
\begin{equation}
V_\mu=\left(
    \begin{array}{ccc}
      \frac{1}{\sqrt{2}}\rho^0+\frac{1}{\sqrt{2}}\omega & \rho^+ & K^{\ast+} \\
      \rho^- & -\frac{1}{\sqrt{2}}\rho^0+\frac{1}{\sqrt{2}}\omega & K^{\ast0} \\
      K^{\ast-} & \bar{K}^{\ast0} & \phi \\
    \end{array}
  \right)_\mu.
\end{equation}

The two-vector-two-pseudoscalar amplitudes can be obtained by expanding the Lagrangian of Eq.~(\ref{Lagrangian}) up to two pseudoscalar meson fields:
\begin{equation}\label{LOV}
\mathcal{L}_{VVPP}=-\frac{1}{4f^2}\textrm{Tr}([V^\mu,\partial^\nu V_\mu][P,\partial_\nu P]),
\end{equation}
which is the so-called Weinberg-Tomozawa interaction for the $VP\rightarrow VP$ process. In the pseudoscalar octet we assume $\eta_8\equiv\eta$. In the vector meson multiplet ideal $\omega_1\raisebox{0.2mm}{--}\omega_8$ mixing is assumed:
\begin{equation}
\phi=\omega_1/\sqrt{3}-\omega_8\sqrt{2/3},\quad \omega=\omega_1\sqrt{2/3}+\omega_8/\sqrt{3}.
\end{equation}
Throughout the work, the following phase convention is used: $|\pi^+\rangle=-|1+1\rangle$,$ |\rho^+\rangle=-|1+1\rangle$, $|K^-\rangle=-|1/2-1/2\rangle$ and $|K^{\ast-}\rangle=-|1/2-1/2\rangle$ , corresponding to $|II_3\rangle$ isospin states.

From the Lagrangian of Eq.~(\ref{LOV}) one obtains the S-wave amplitude:
\begin{equation}
V_{ij}(s)=-\frac{\epsilon\cdot\epsilon'}{8f^2}C_{ij}[3s-(M^2+m^2+M'^2+m'^2)-\frac{1}{s}(M^2-m^2)(M'^2-m'^2)],
\end{equation}
where the $\epsilon$ ($\epsilon'$) stands for the polarization four-vector of the incoming (outgoing) vector meson. The masses $M$ ($M'$), $m$ ($m'$) correspond to the initial (final) vector mesons and initial (final) pseudoscalar mesons, respectively. The indices $i$ and $j$ represent the initial and final $VP$ states, respectively. The $C_{ij}$ coefficients for $VP$ coupled channels  in the $(S,I)=(1,1/2)$, $(0,0)$, and $(0,1)$ sectors can be found in Ref.~\cite{Roca:2005nm},
where $S$ denotes strangeness and $I$ represents isospin.

\subsection{NLO chiral potentials}
The chiral Lagrangians relevant to the present study  up to chiral order $\mathcal{O}(p^2)$ and with just one trace in flavor space
(leading order in the large $N_c$ expansion) has the following form~\cite{Prades:1993ys}:
\begin{equation}\label{Eq:NLOL}
\begin{split}
\mathcal{L}_{VV}^{(2)}=&\frac{1}{2}\delta_1\textrm{TR}(V_\mu V^\mu u_\nu u^\nu)+\frac{1}{2}\delta_2\textrm{Tr}(V_\mu u_\nu V^\mu u^\nu)\\ &+\frac{1}{2}\delta_3\textrm{Tr}(V_\mu V_\nu u^\mu u^\nu)+\frac{1}{2}\delta_4\textrm{Tr}(V_\mu V_\nu u^\nu u^\mu)\\ &+\frac{1}{2}\delta_5\textrm{Tr}(V_\mu u^\mu V_\nu u^\nu+V_\mu u_\nu V^\nu u^\mu)\\ &+\frac{1}{2}\kappa_V\textrm{Tr}(V_\mu V^\mu \chi_{+})\end{split},
\end{equation}
where $
u_\mu=i\{u^{\dag}\partial_\mu u-u\partial_\mu u^\dag\} $
and  $\chi_{\pm}=u^\dag\chi u^\dag\pm u\chi^\dag u $
with $\chi=$diag$(m^2_\pi,m^2_\pi,2m^2_K-m^2_\pi)$. 

In Eq.~(\ref{Eq:NLOL}) the low-energy constant (LEC)  $\kappa_V$ is readily determined from the $K^\ast\raisebox{0.2mm}{--}\rho$ mass splitting:
\begin{equation}
\kappa_V=\frac{M^2_{K^\ast}-M^2_\rho}{2(m^2_K-m_\pi^2)}.
\end{equation}

The NLO chiral potentials, after projection into S-wave, are
\begin{equation}\label{eq:nloV}
V^\mathrm{NLO}=\left[\frac{\delta_1D^{(1)}_{ij}+\delta_2 D^{(2)}_{ij}}{f^2} E_{P}E_{P'}+\frac{\kappa_V}{2 f^2} D^{(3)}_{ij}\right]\epsilon\cdot\epsilon',
\end{equation}
where $E_P$ and $E_{P'}$ are the energy of the initial and final pseudoscalar mesons, and
$D^{(1,2,3)}_{ij}$ are the corresponding Clebsch-Gordan coefficients and are tabulated in the appendix.

\subsection{Unitarization}

The unitarized amplitude obtained from the Bethe-Salpeter approach~\cite{Roca:2005nm} is:
\begin{equation}
T=[1+V\hat{G}]^{-1}(-V)\vec{\epsilon}\cdot\vec{\epsilon}',
\end{equation}
where $\vec{\epsilon}$ ($\vec{\epsilon}'$) is the spatial polarization vector of the initial (final) vector meson, and $\hat{G}=G(1+\frac{1}{3}\frac{q_l^2}{M_l^2})$ is a diagonal matrix with the $l\raisebox{0.2mm}{--}$th element, $G_l$ being the two-meson loop function containing a vector and a pseudoscalar meson:
\begin{equation}\label{eq:loop}
G_l(\sqrt{s})=i\int\frac{d^4q}{(2\pi)^4}\frac{1}{(P-q)^2-M_l^2+i\epsilon}\frac{1}{q^2-m_l^2+i\epsilon}
\end{equation}
with $P$ the total incident momentum, which  is $(\sqrt{s},0,0,0)$ in the center of mass frame. In the dimensional regularization scheme the loop function reads as
\begin{equation}\label{eq:loopf}\small
\begin{split}
G_l(\sqrt{s})=\frac{1}{16\pi^2}\{a(\mu)&+\textrm{ln}\frac{M_l^2}{\mu^2}+\frac{m_l^2-M_l^2+s}{2s}\textrm{ln}\frac{m_l^2}{M_l^2}\\ +\frac{q_l}{\sqrt{s}}&[\textrm{ln}(s-(M_l^2-m_l^2)+2q_l\sqrt{s})+\textrm{ln}(s+(M_l^2-m_l^2)+2q_l\sqrt{s})\\ &-\textrm{ln}(s-(M_l^2-m_l^2)-2q_l\sqrt{s})-\textrm{ln}(s+(M_l^2-m_l^2)-2q_l\sqrt{s})-2\pi i]\},
\end{split}
\end{equation}
where $\mu$ is the dimensional regularization scale. Changes in the scale are reabsorbed in the subtraction constant $a(\mu)$, so that the results remain scale independent. In Eq.~(\ref{eq:loopf}), $q_l$ denotes the three-momentum of the vector or pseudoscalar meson in the center of mass frame and is given by:
\begin{equation}
q_l=\frac{1}{2\sqrt{s}}\sqrt{[s-(M_l+m_l)^2][s-(M_l-m_l)^2]},
\end{equation}
where $M_l$ and $m_l$ are the masses of the vector and pseudoscalar mesons, respectively.

In Ref.~\cite{ Altenbuchinger:2013vwa}, the power-counting issue related to the regularization of the loop function $G_l$ [Eq.~(\ref{eq:loop})] was discussed.
It was found that the usual treatment as done in Eq.~(\ref{eq:loopf}) is reasonable by assuming that $a(\mu)$ is in fact a $\mathcal{O}(p)$ quantity in order to ensure that $G_l$ is of $\mathcal{O}(p)$.  The on-shell approximation adopted in the present work is carefully examined in Ref.~\cite{Altenbuchinger:2013gaa}, where
special attention is paid to its influence on the light-quark mass dependence of the scattering lengths and the pole positions. It is shown that results obtained from the on-shell approximation can be trusted at least for quantities not far away from their respective threshold.
\section{Results and discussion}
In this section, we study the impact of the NLO chiral potentials on the dynamically generated axial-vector mesons. In anticipation of future LQCD simulations, we further calculate the scattering lengths and the pole positions as  functions of the light-quark masses. As in Ref.~\cite{Roca:2005nm}, we fix $f=92$ MeV, $a=-1.85$, and $\mu=900$ MeV. The isospin averaged physical pseudoscalar and vector meson masses are taken from Ref.~\cite{Beringer:1900zz}.

\subsection{Impact of NLO potentials}
At $\mathcal{O}(p^2)$, there are six additional LECs. As mentioned earlier, the LEC $\kappa_V$ can be fixed from the $\rho$-$K^*$ mass splitting. In addition, the contributions proportional to the LECs $\delta_{3,4,5}$ are  suppressed compared to those proportional to $\delta_1$ and $\delta_2$. Indeed, the $\delta_3$, $\delta_4$, and $\delta_5$ terms vanish in the $M_V\rightarrow\infty$ limit, where $M_V$ is a generic vector meson mass.\footnote{For a related discussion in the heavy-light meson sector, see Ref.~\cite{Altenbuchinger:2013vwa}.}  Therefore, we only take into account the terms proportional to  $\delta_1$ and $\delta_2$ and study their effects on the dynamically generated axial-vector mesons.

Since our primary purpose is to check whether the conclusions of Refs.~\cite{Lutz:2003fm,Roca:2005nm} are stable against the inclusion of the NLO chiral potentials as kernels in the Bethe-Salpeter equation, we decide to vary the LECs $\delta_1$ and $\delta_2$ between $-1$ and $1$, assuming that they are of natural size, and then check whether the axial-vector mesons can still be dynamically generated. Indeed, we find that the dynamical generation of the axial vector mesons does not depend sensitively on the
values of the LECs $\delta_1$ and $\delta_2$.  

\begin{figure}[t]
\includegraphics[width=1\textwidth]{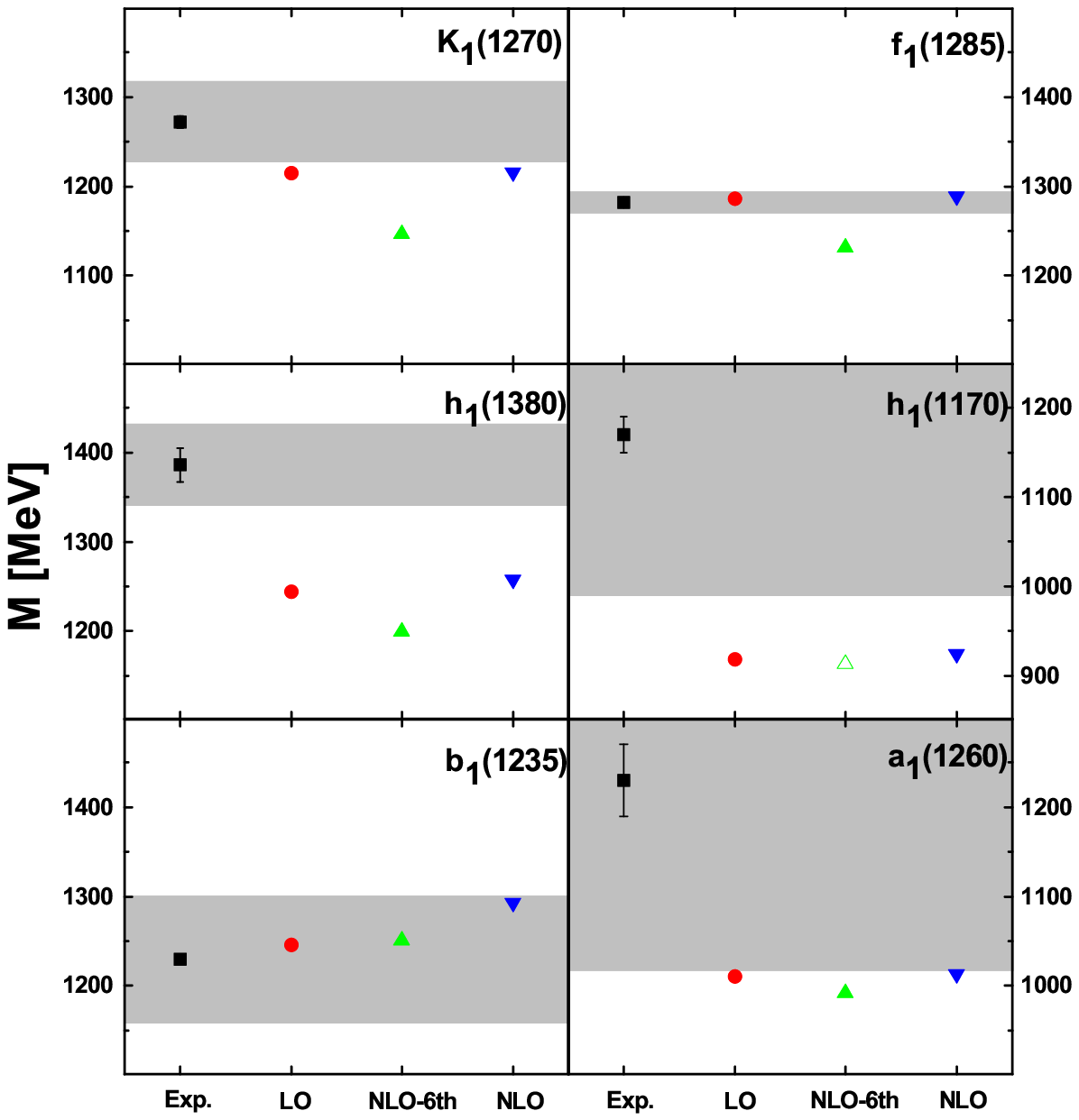}
\caption{The experimental axial-vector masses~\cite{Beringer:1900zz} in comparison with the corresponding UChPT results. The LO, NLO-6th, and NLO numbers denote the masses obtained from the
LO kernel, the NLO kernel containing only the $\kappa_V$ term, and the complete NLO kernel. The error bars denote the uncertainties of the experimental masses while the shaded bands indicate the experimental widths. For the $K_1(1260)$, only the high-energy pole, coupling mainly to $\rho K$, is shown.}
\label{Fig:nloeffects}
\end{figure}

In Fig.~\ref{Fig:nloeffects}, we compare the experimental masses of the axial vector mesons with those obtained from the UChPT with the LO kernel and the NLO kernel. The LECs $\delta_1$ and $\delta_2$ are fixed at 1.0 and 0.2, respectively. These values are chosen to yield an overall good agreement between the experimental and theoretical masses. It is clear that the mass degeneracy breaking term in the NLO kernel alone deteriorates the LO results. The inclusion of the $\delta_1$ and $\delta_2$ terms can improve the description, but not much. The overall agreement with the experimental masses is similar to the LO results.

In Tables~\ref{tab:pole1}, ~\ref{tab:pole2}, and~\ref{tab:pole3}, we tabulate the pole positions and their respective couplings to each channel. 
 The couplings are obtained from the residues of the $T$ matrix elements at the pole position $s_0$, i.e.,
\begin{equation}\label{eq:residue}
T_{ij}=\frac{g_i g_j}{s-s_0}.
\end{equation}
It can be seen that the couplings obtained with the LO and NLO kernels are similar to each other.

\begin{table}[htbp]
  \caption{Pole positions and couplings in the  $(S,I)=(1,1/2)$ channel}\label{tab:pole1}
  \centering
  \begin{tabular}{ccc|cc}
    \hline\hline
     & \multicolumn{4}{c}{$K_1(1270)$} \\
    \hline
     & LO & NLO & LO & NLO \\
    \cline{2-5}
    $\sqrt{s}$ & $1111-i64$ & $1119-i68$ & $1215-i4$ & $1216-i5$ \\
    \hline
    $\phi K$ & $1584-i873$ & $2148-i593$ & $1095-i401$ & $1867-i505$ \\
    $\omega K$ & $-1858+i649$ & $-2434+i1063$ & $-1031+i375$ & $-462+i605$ \\
    $\rho K$ & $-1522+i1155$ & $-1121+i1311$ & $5268+i298$ & $5365+i317$ \\
    $K^\ast\eta$ & $27+i156$ & $-107+i382$ & $3454-i93$ & $4964-i147$ \\
    $K^\ast\pi$ & $4186-i2098$ & $4352-i2075$ & $339-i984$ & $391-i1197$ \\
    \hline\hline
  \end{tabular}
\end{table}

\begin{table}[htbp]
  \caption{Pole positions and couplings in the $(S,I)=(0,0)$ channel}\label{tab:pole2}
  \centering
  \begin{tabular}{ccccccc}
    \hline\hline
     & \multicolumn{2}{c}{$h_1(1170)$} & \multicolumn{2}{c}{$h_1(1380)$} & \multicolumn{2}{c}{$f_1(1285)$} \\
    \hline
     & LO & NLO & LO & NLO & LO & NLO \\
    \cline{2-7}
    $\sqrt{s}$ & $918-i17$ & $925-i29$ & $1244-i7$ & $1257-i0$ & $1286-i0$ & $1289-i0$ \\
    \hline
    $\frac{1}{\sqrt{2}}(\bar{K}^\ast K+K^\ast \bar{K})$ & $...$ & $...$ & $...$ & $...$ & $7219+i0$ & $7884+i0$ \\
    $\phi\eta$ & $-46+i13$ & $69-i102$ & $-3309+i47$ & $-5963-i38$ & $...$ & $...$ \\
    $\omega\eta$ & $-24+i28$ & $711-i427$ & $3019-i22$ & $2642-i47$ & $...$ & $...$ \\
    $\rho\pi$ & $3452-i1681$ & $3576-i1909$ & $650-i961$ & $134-i233$ & $...$ & $...$ \\
    $\frac{1}{\sqrt{2}}(\bar{K}^\ast K-K^\ast \bar{K})$ & $-784+i499$ & $-1488+i757$ & $6137+i183$ & $6435+i35$ & $...$ & $...$ \\
    \hline
  \end{tabular}
\end{table}

\begin{table}[htbp]
  \caption{Pole positions and couplings in the $(S,I)=(0,1)$ channel}\label{tab:pole3}
  \centering
  \begin{tabular}{ccccc}
    \hline\hline
     & \multicolumn{2}{c}{$a_1(1260)$} & \multicolumn{2}{c}{$b_1(1235)$} \\
    \hline
     & LO & NLO & LO & NLO \\
    \cline{2-5}
    $\sqrt{s}$ & $1011-i84$ & $1013-i106$ & $1246-i27$ & $1293-i22$ \\
    \hline
    $\frac{1}{\sqrt{2}}(\bar{K}^\ast K+K^\ast \bar{K})$ & $...$ & $...$ & $6159-i75$ & $5501-i38$ \\
    $\phi\pi$ & $...$ & $...$ & $2085-i384$ & $2898-i480$ \\
    $\omega\pi$ & $...$ & $...$ & $-1867+i299$ & $-111-i77$ \\
    $\rho\eta$ & $...$ & $...$ & $-3040+i496$ & $-2458+i345$ \\
    $\rho\pi$ & $3794-i2328$ & $3931-i2794$ & $...$ & $...$ \\
    $\frac{1}{\sqrt{2}}(\bar{K}^\ast K-K^\ast \bar{K})$ & $-1875+i1485$ & $-2649+1928$ & $...$ & $...$ \\
    \hline\hline
  \end{tabular}
\end{table}

\subsection{Scattering lengths of pseudoscalar mesons and vector mesons}
Scattering lengths provide vital information on the nature of hadron-hadron interactions at threshold and can show a first hint on whether
the interaction is strong enough to form shallow bound states and resonances.  On the lattice, the scattering lengths can be easily calculated and their light-quark mass dependences prove to be of great value to unravel the nature of the underlying interactions. Recently, several studies have shown that by fitting to the
LQCD data on the scattering lengths of the pseudoscalar mesons off the $D$ mesons, the UChPT can generate dynamically the $D_{s_0}^*(2317)$ without a priori assumption of its existence~\cite{Altenbuchinger:2013vwa,Liu:2012zya}. Such studies demonstrate that a lot can be accomplished by combining first-principles LQCD simulations with effective field theories, such as the UChPT.

The scattering lengths of channel $i$ with strangeness $S$ and isospin $I$ 
are related to the diagonal $T$-matrix elements $T_{ii}$ in the following way:
\begin{equation}
a_i^{(S,I)}=-\frac{1}{8\pi (M_1+m_2)}T_{ii}^{(S,I)}(s=(M_1+m_2)^2),
\end{equation}
where $M_1$ and $m_2$ are the masses of the vector meson and the pseudoscalar meson of channel $i$.
Since our aim is to study the light-quark mass dependence of the scattering lengths, we need
to know the light-quark mass dependence of the pseudoscalar mesons and the vector mesons.  One can turn to LQCD simulations for the relevant information. 
Here we use the recent $n_f=2+1$ results of the PACS-CS Collaboration performed with the nonperturbatively $\mathcal{O}(a)$-improved Wilson quark action and the Iwasaki gauge
action~\cite{Aoki:2008sm}.   From the simulated pion and kaon masses,
one obtains the following relation~\cite{Ren:2012aj}:
\begin{equation}\label{eq:mK}
  m_K^2=a+b m_\pi^2,
\end{equation}
where $a=0.291751\,\mathrm{GeV}^2$ and $b= 0.670652$.
In our present work, we use the LO ChPT to relate the eta meson mass to those of the kaon and the pion, i.e.,
\begin{equation}\label{eq:meta}
m_\eta^2=\frac{4 m_K^2-m_\pi^2}{3}.
\end{equation}

The vector meson masses up to $\mathcal{O}(p^2)$ can be calculated from the following Lagrangian:
\begin{equation}\label{the Lagrangians of mass terms}
  \mathcal{L}_{VV}=\frac{1}{2}M_0^2\textrm{Tr}(V^\mu V_\mu)+\frac{\lambda_m}{2} \textrm{Tr}(V_\mu V^\mu \chi_+)+\frac{\lambda_0}{4} \textrm{Tr}(V_\mu V^\mu)\textrm{Tr}(\chi_+),
\end{equation}
yielding
\begin{eqnarray}\label{eq:mV}
  M_\rho^2 &=&M_0^2+2\lambda_m m_\pi^2+\lambda_0(2m_K^2+m_\pi^2),\nonumber\\
  M_{K^*}^2&=&M_0^2+2\lambda_m m_K^2+\lambda_0(2m_K^2+m_\pi^2),\nonumber\\
  M_{\phi}^2&=&M_0^2+4\lambda_m m_K^2-2\lambda_m m_\pi^2+\lambda_0(2m_K^2+m_\pi^2),\nonumber\\
  M_\omega^2&=&M_0^2+2\lambda_m m_\pi^2+\lambda_0(2m_K^2+m_\pi^2).
\end{eqnarray}
Performing a least-of-squares fit to the PACS-CS data~\cite{Aoki:2008sm}, the chiral limit vector meson mass $M_0$, and the LECs $\lambda_m$ and $\lambda_0$ are determined to be
\begin{equation}
  M_0=0.711964\,\mathrm{GeV}, \lambda_m=0.48901, \lambda_0=0.126032,
\end{equation}
with a $\chi^2/\mathrm{d.o.f.}=0.723$.\footnote{Fitting the LQCD vector meson masses together with their experimental
counterpart yields very similar results and has no appreciable effect on the results shown below.}  We notice that the $\lambda_m$ determined from fitting the PACS-CS data is only slightly different
from the $\kappa_V$ determined using the physical $K^*$-$\rho$ mass difference, which is $0.434$.

\begin{figure}[t]
\includegraphics[width=1\textwidth]{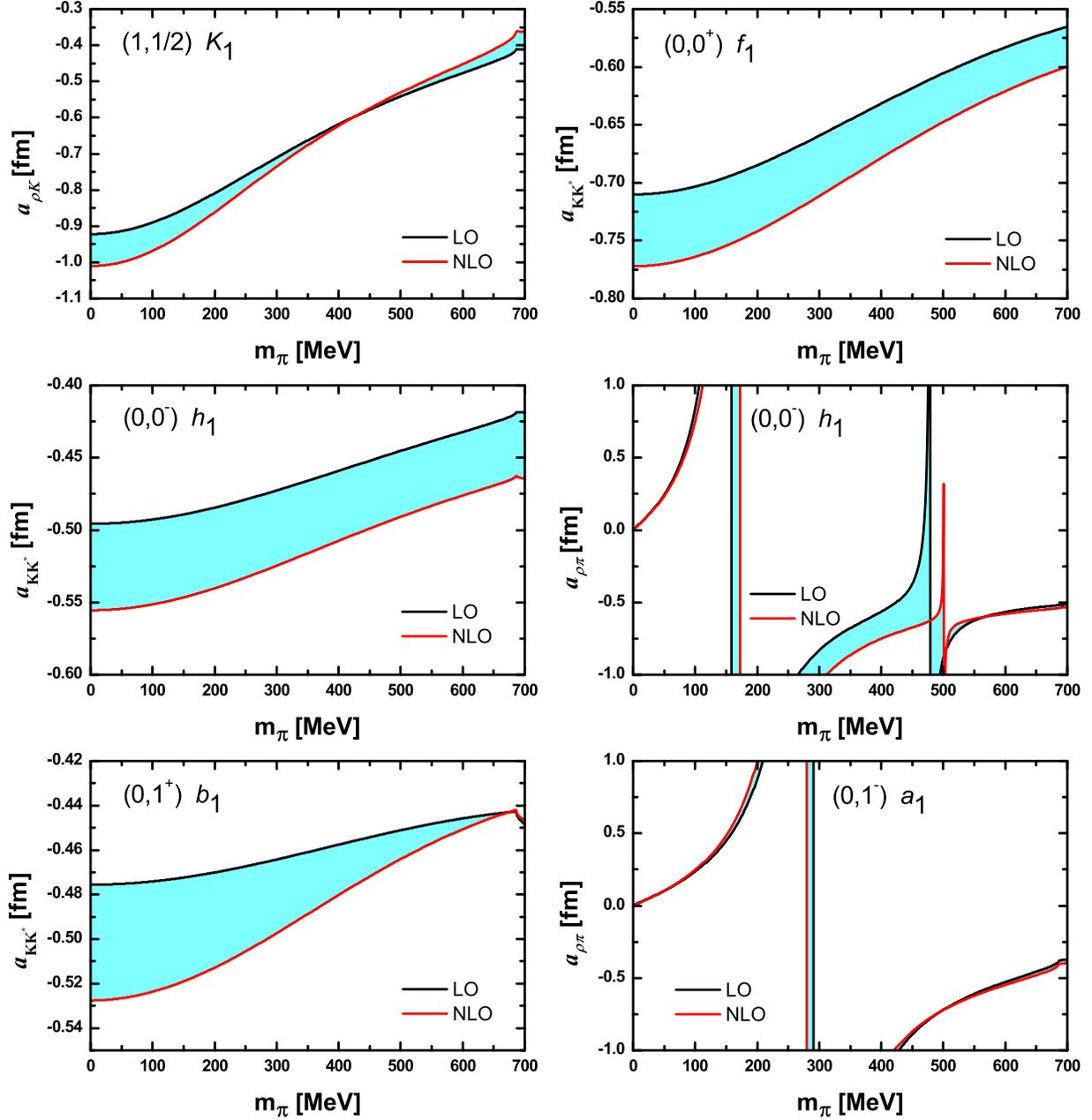}
\caption{Scattering lengths in the dominant channels of each isospin, strangeness, and G parity $(S,I^{(G)})$  sector.}
\label{Fig:SL}
\end{figure}

In Fig. \ref{Fig:SL}, we show the scattering length of the dominant channels (determined by the couplings tabulated in Tables ~\ref{tab:pole1}, ~\ref{tab:pole2}, and~\ref{tab:pole3}) in each isospin, strangeness, and G parity sector as a function of the pion mass,
with the strangeness quark mass   fixed at its physical value using the LO ChPT and the vector meson masses given by Eqs.~(\ref{eq:mV}).
The difference between the results obtained with the LO kernel and the NLO kernel indicates inherent theoretical uncertainty.  Interestingly, we notice that the $a_{\rho\pi}$ shows some ``threshold'' effects  in the $h_1$ channel 
and in the $a_1$ channel. Such threshold effects can be easily understood from Eq.~(\ref{eq:residue}) and Fig.~\ref{fig:pole}. When one varies the light quark masses,  as routinely done in LQCD simulations, 
a bound state for a certain channel can become a resonance for that chanel, i.e., the trajectory of the threshold crosses the pole trajectory, as shown in Fig.~\ref{fig:pole}.  If this happens, one can immediately see that for $s$ approaching $s_0$ from below, $a\rightarrow \infty$, while just above $s_0$, $a\rightarrow -\infty$.  In performing LQCD simulations of scattering lengths, one needs to be aware of such a scenario.

In principle, one can compare directly the scattering lengths  at unphysical light-quark masses  with those obtained from a LQCD simulation. For instance, in Ref.~\cite{Lang:2014tia},
the scattering length of $\rho$ and $\pi$ mesons, $a_{\rho\pi}$, is found to be $0.62(28)$ at a pion mass of $266$ MeV. In our case, the $a_{\rho\pi}$ is much larger at the order of 10, as can be seen from Fig.~\ref{Fig:SL}. Such a comparison needs to 
be taken with caution, however. First, the LQCD simulations are performed with $n_f=2$ configurations. But more importantly, the threshold effects can make such a comparison around the
``threshold'' region unreliable, because  these effects depend sensitively on the particular light-quark mass dependence pattern of the masses of the building blocks, in the present work, those of the pseudoscalar and vector mesons. 
This may or may not be realized for a particular LQCD  simulation. This being said, for resonances consisting of large hadron-hadron components, one may need to study the related scattering lengths carefully in order to not misinterpret the data in case that such a scenario happens.

\subsection{Light-quark mass dependence of the pole positions}
\begin{figure}[t]
\includegraphics[width=1\textwidth]{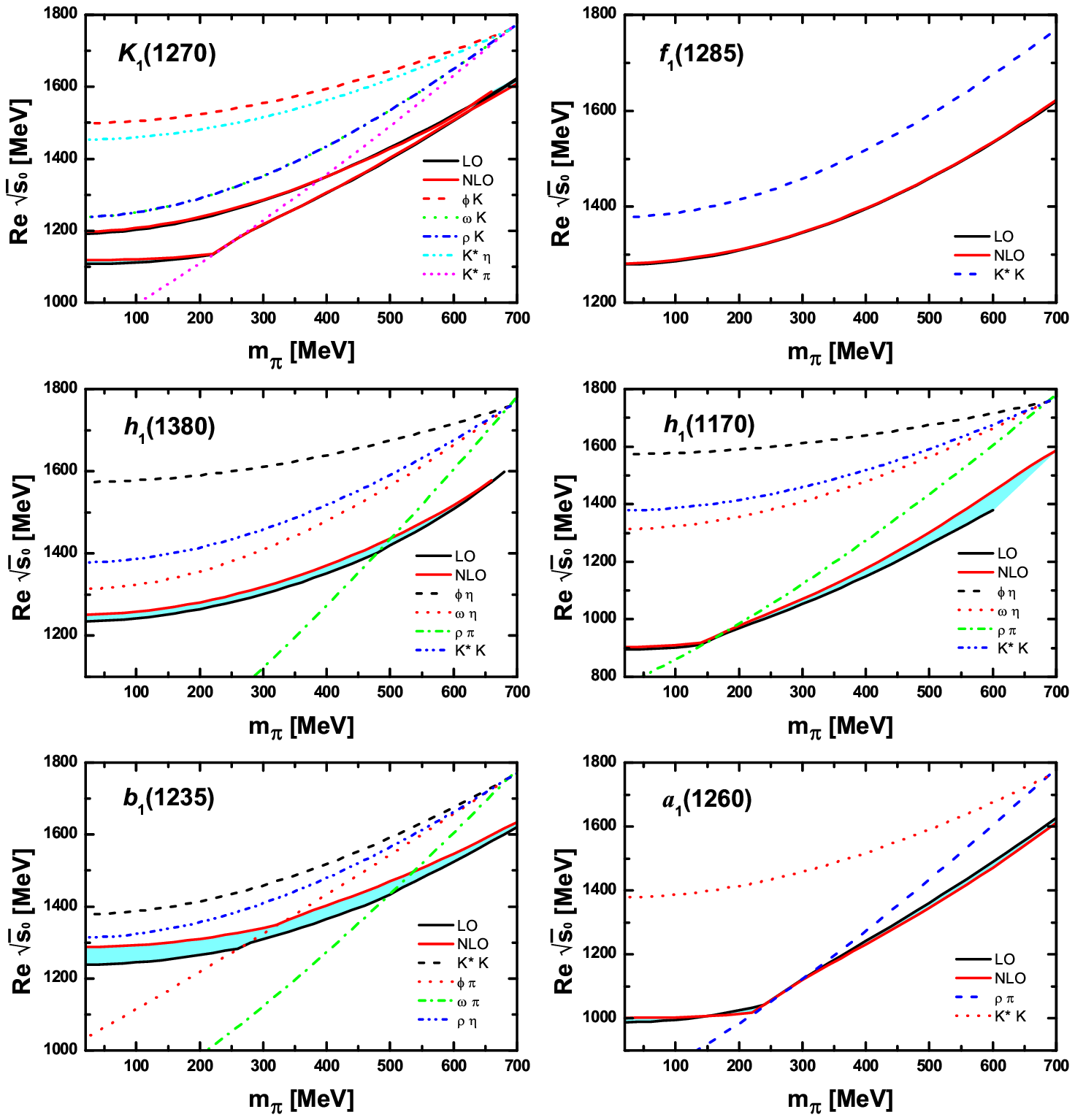}
\caption{Real part of the pole positions, $\mathrm{Re} \sqrt{s_0}$,  as a function of $m_\pi$ .  The dashed lines indicate the thresholds of the corresponding coupled channels, respectively.}
\label{fig:pole}
\end{figure}
\begin{figure}[t]
\includegraphics[width=1\textwidth]{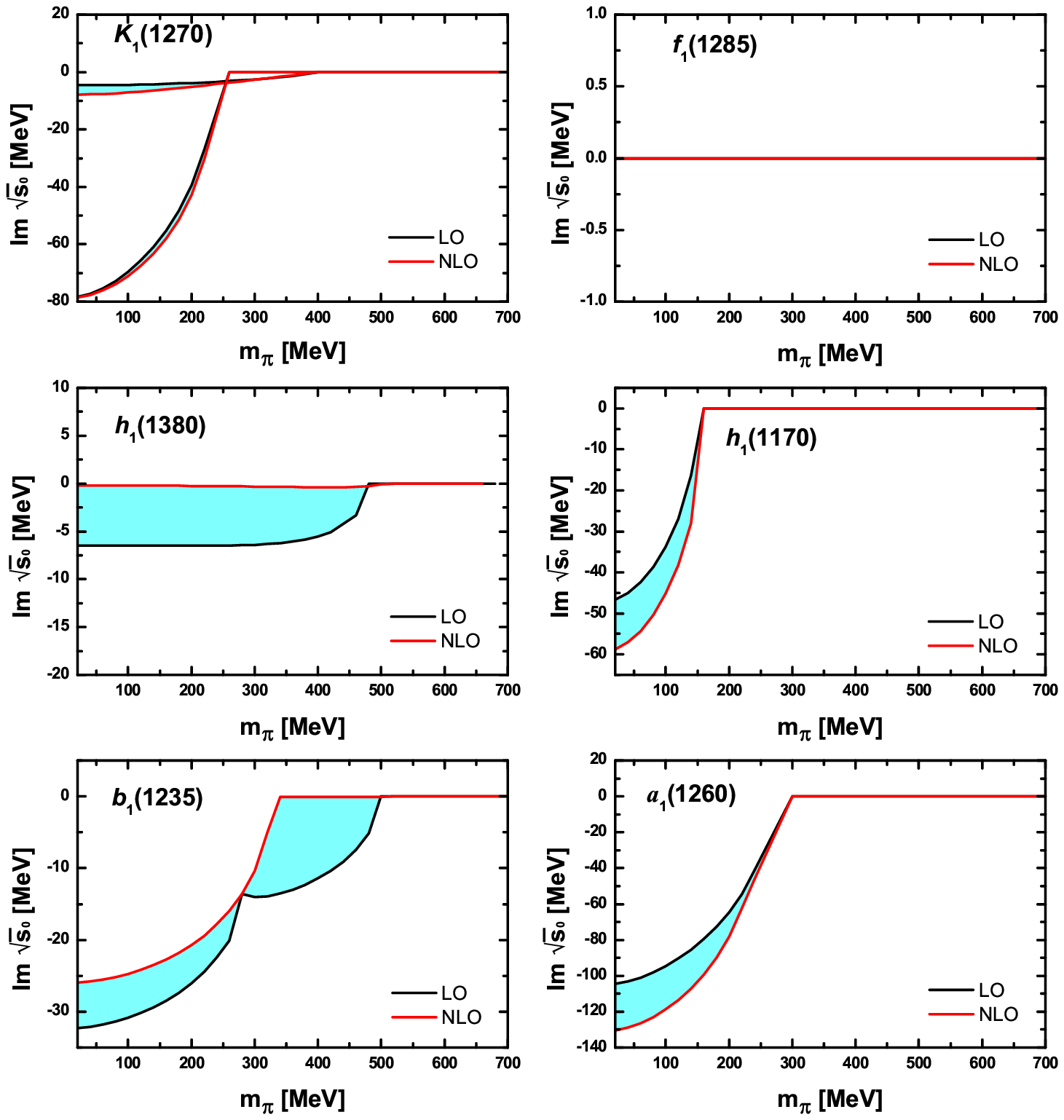}
\caption{Imaginary part of the pole positions, $\mathrm{Im} \sqrt{s_0}$,  as a function of $m_\pi$ .}
\label{fig:poleim}
\end{figure}
Because  LQCD simulations are generally performed with larger than physical light-quark masses, the so-obtained resonance properties are
not those of the physical world. It is of great importance to have a reliable approach to perform the necessary chiral extrapolations. In addition, the light-quark mass dependence of the pole positions can yield valuable information on the nature of the resonances, as have been argued in a number of previous studies~\cite{ Cleven:2010aw,Hanhart:2008mx}. 

In Fig.~\ref{fig:pole}, the real parts of the pole positions corresponding to the two $K_1(1270)$ states, the $a_1(1260)$, the $b_1(1235)$,
the $f_1(1285)$, the $h_1(1170)$, and the $h_1(1380)$ are shown as functions of the pion mass. It
is interesting to note that all the states exist in the range of $m_\pi$ till 700 MeV and thus LQCD simulations should have no problem in identifying them
even at unphysical light-quark masses. In addition, the two $K_1(1270)$ states persist and remain well separated, thus allowing for the possibility of being identified on the lattice. 
We notice that recently a strategy has been proposed to extract the two-pole structure of the $\Lambda(1405)$ from LQCD simulations in a finite box\cite{MartinezTorres:2012yi}. Such a strategy may also be employed to
study the two-pole structure of the $K_1(1270)$ on the lattice.

In Fig.~\ref{fig:poleim},  the imaginary parts of the pole positions are shown. It is clear that as the light-quark mass becomes large, these axial-vector mesons
become bound. While as the pion mass decreases, even still two times larger than its physical value, they already become a resonance. Therefore, to obtain reliable results, LQCD simulations at small light-quark masses may need to take into account coupled channel effects and final state interactions explicitly, among others. Such studies are very demanding both numerically and theoretically. Nevertheless encouraging attempts have been seen recently.

Finally, we should mention that the vector mesons have finite widths, particularly, those of the $K^*$ and the $\rho$. We have checked that including the widths as suggested in Ref.~\cite{Geng:2006yb} does not change qualitatively the
obtained pole positions and couplings. On the other hand, we have no information on how the widths change with varying light quark masses and therefore cannot incorporate the widths into our studies of the scattering lengths and the pole positions without making further assumptions. We will leave such a study to a future work.
\section{Summary}
In this work, we have studied the effects of the next-to-leading order chiral potential on the dynamically generated axial vector mesons. We found that the inclusion of the higher-order kernel does not change the results obtained with the leading-order kernel in any significant way, and thus lend more confidence on the dynamical picture. In anticipation of future LQCD simulations of these resonances, utilizing the PACS-CS simulations of the pseudoscalar and vector mesons masses, we have calculated the scattering lengths and the pole positions as functions of the pion mass.  These results, when contrasted with future LQCD simulations, may provide a clue on the true nature of the axial-vector mesons.

We have shown that when the masses of the pseudoscalar and the vector mesons vary with the light-quark masses, one may observe threshold like effects in the related scattering lengths. Future LQCD simulations may need to be carefully analyzed and should not be misinterpreted if such a scenario is realized.

\section{Acknowledgements}

We thank  Dr. Eulogio Oset for a careful reading of this manuscript. X.-L.R acknowledges support from the Innovation Foundation of Beihang University for Ph.D. graduates.  This work is supported in part  by the National Natural Science Foundation of China under Grant Nos. 11375024 and 11205011,
and the New Century Excellent Talents in University Program of Ministry of Education of China under
Grant No. NCET-10-0029.

\section{Appendix}
In this section, we tabulate the Clebsch-Gordan coefficients  $D_{ij}^{(1,2,3)}$ appearing In the NLO chiral potentials [Eq.~(\ref{eq:nloV})] for $(S,I)=(1,1/2)$, $(0,0)$, and $(0,1)$.

\begin{table}[htbp]
  \caption{ $D^{(1)}_{ij}$ of Eq.~(\ref{eq:nloV}) for $(S,I)=(1,\frac{1}{2})$.} \label{coefficients 1st term 1 1/2}
  \centering
  \begin{tabular}{p{40pt}p{40pt}<{\centering}p{40pt}<{\centering}p{40pt}<{\centering}p{40pt}<{\centering}p{40pt}<{\centering}}
  \hline\hline
   & $\phi K$ & $\omega K$ & $\rho K$ & $K^\ast\eta$ & $K^\ast\pi$ \\
  \hline
  $\phi K$ & $2$ & $0$ & $0$ & $-\frac{1}{\sqrt{6}}$ & $\sqrt{\frac{3}{2}}$ \\
  $\omega K$ & $0$ & $1$ & $-\sqrt{3}$ & $-\frac{1}{2\sqrt{3}}$ & $\frac{\sqrt{3}}{2}$ \\
  $\rho K$ & $0$ & $-\sqrt{3}$ & $1$ & $\frac{1}{2}$ & $\frac{1}{2}$ \\
  $K^\ast\eta$ & $-\frac{1}{\sqrt{6}}$ & $-\frac{1}{2\sqrt{3}}$ & $\frac{1}{2}$ & $\frac{5}{3}$ & $1$ \\
  $K^\ast\pi$ & $\sqrt{\frac{3}{2}}$ & $\frac{\sqrt{3}}{2}$ & $\frac{1}{2}$ & $1$ & $1$ \\
  \hline\hline
  \end{tabular}
\end{table}

\begin{table}[htbp]
  \caption{ $D^{(1)}_{ij}$ of  Eq.~(\ref{eq:nloV}) for  $(S,I)=(0,0)$.} \label{coefficients 1st term 0 0}
  \centering
  \begin{tabular}{p{20pt}p{90pt}<{\centering}p{90pt}<{\centering}p{30pt}<{\centering}p{30pt}<{\centering}p{30pt}<{\centering}p{90pt}<{\centering}}
  \hline\hline
  $G$ & & $\frac{1}{\sqrt{2}}(\bar{K}^\ast K+K^\ast \bar{K})$ & $\phi\eta$ & $\omega\eta$ & $\rho\pi$ & $\frac{1}{\sqrt{2}}(\bar{K}^\ast K-K^\ast \bar{K})$ \\
  \hline
  $+$ & $\frac{1}{\sqrt{2}}(\bar{K}^\ast K+K^\ast \bar{K})$ & $3$ & $0$ & $0$ & $0$ & $0$ \\
  $-$ & $\phi\eta$ & $0$ & $\frac{8}{3}$ & $0$ & $0$ & $-\sqrt{\frac{2}{3}}$ \\
  $-$ & $\omega\eta$ & $0$ & $0$ & $\frac{2}{3}$ & $-2$ & $-\frac{1}{\sqrt{3}}$ \\
  $-$ & $\rho\pi$ & $0$ & $0$ & $-2$ & $2$ & $-\sqrt{3}$ \\
  $-$ & $\frac{1}{\sqrt{2}}(\bar{K}^\ast K-K^\ast \bar{K})$ & $0$ & $-\sqrt{\frac{2}{3}}$ & $-\frac{1}{\sqrt{3}}$ & $-\sqrt{3}$ & $3$ \\
  \hline
  \hline
  \end{tabular}
\end{table}

\begin{table}[htbp]
  \caption{ $D^{(1)}_{ij}$ of Eq.~(\ref{eq:nloV}) for $(S,I)=(0,1)$.} \label{coefficients 1st term 0 1}
  \centering
  \begin{tabular}{p{20pt}p{90pt}<{\centering}p{90pt}<{\centering}p{30pt}<{\centering}p{30pt}<{\centering}p{30pt}<{\centering}p{30pt}<{\centering}p{90pt}<{\centering}}
  \hline\hline
  $G$ & & $\frac{1}{\sqrt{2}}(\bar{K}^\ast K+K^\ast \bar{K})$ & $\phi\pi$ & $\omega\pi$ & $\rho\eta$ & $\rho\pi$ & $\frac{1}{\sqrt{2}}(\bar{K}^\ast K-K^\ast \bar{K})$ \\
  \hline
  $+$ & $\frac{1}{\sqrt{2}}(\bar{K}^\ast K+K^\ast \bar{K})$ & $1$ & $-\sqrt{2}$ & $-1$ & $\frac{1}{\sqrt{3}}$ & $0$ & $0$ \\
  $+$ & $\phi\pi$ & $-\sqrt{2}$ & $0$ & $0$ & $0$ & $0$ & $0$ \\
  $+$ & $\omega\pi$ & $-1$ & $0$ & $2$ & $\frac{2}{\sqrt{3}}$ & $0$ & $0$ \\
  $+$ & $\rho\eta$ & $\frac{1}{\sqrt{3}}$ & $0$ & $\frac{2}{\sqrt{3}}$ & $\frac{2}{3}$ & $0$ & $0$ \\
  $-$ & $\rho\pi$ & $0$ & $0$ & $0$ & $0$ & $2$ & $-\sqrt{2}$ \\
  $-$ & $\frac{1}{\sqrt{2}}(\bar{K}^\ast K-K^\ast \bar{K})$ & $0$ & $0$ & $0$ & $0$ & $-\sqrt{2}$ & $1$ \\
  \hline
  \hline
  \end{tabular}
\end{table}

\begin{table}[htbp]
  \caption{ $D^{(2)}_{ij}$ of Eq.~(\ref{eq:nloV}) for $(S,I)=(1, \frac{1}{2})$.} \label{coefficients 2nd term 1 1/2}
  \centering
  \begin{tabular}{p{40pt}p{40pt}<{\centering}p{40pt}<{\centering}p{40pt}<{\centering}p{40pt}<{\centering}p{40pt}<{\centering}}
  \hline\hline
   & $\phi K$ & $\omega K$ & $\rho K$ & $K^\ast\eta$ & $K^\ast\pi$ \\
  \hline
  $\phi K$ & $0$ & $\sqrt{2}$ & $-\sqrt{6}$ & $-2\sqrt{\frac{2}{3}}$ & $0$ \\
  $\omega K$ & $\sqrt{2}$ & $0$ & $0$ & $\frac{1}{\sqrt{3}}$ & $\sqrt{3}$ \\
  $\rho K$ & $-\sqrt{6}$ & $0$ & $0$ & $-1$ & $-3$ \\
  $K^\ast\eta$ & $-2\sqrt{\frac{2}{3}}$ & $\frac{1}{\sqrt{3}}$ & $-1$ & $-\frac{4}{3}$ & $-2$ \\
  $K^\ast\pi$ & $0$ & $\sqrt{3}$ & $-3$ & $-2$ & $0$ \\
  \hline\hline
  \end{tabular}
\end{table}

\begin{table}[htbp]
  \caption{ $D^{(2)}_{ij}$ of Eq.~(\ref{eq:nloV}) for $(S,I)=(0,0)$.} \label{coefficients 2nd term 0 0}
  \centering
  \begin{tabular}{p{20pt}p{90pt}<{\centering}p{90pt}<{\centering}p{30pt}<{\centering}p{30pt}<{\centering}p{30pt}<{\centering}p{90pt}<{\centering}}
  \hline\hline
  $G$ & & $\frac{1}{\sqrt{2}}(\bar{K}^\ast K+K^\ast \bar{K})$ & $\phi\eta$ & $\omega\eta$ & $\rho\pi$ & $\frac{1}{\sqrt{2}}(\bar{K}^\ast K-K^\ast \bar{K})$ \\
  \hline
  $+$ & $\frac{1}{\sqrt{2}}(\bar{K}^\ast K+K^\ast \bar{K})$ & $-6$ & $0$ & $0$ & $0$ & $0$ \\
  $-$ & $\phi\eta$ & $0$ & $\frac{8}{3}$ & $0$ & $0$ & $-4\sqrt{\frac{2}{3}}$ \\
  $-$ & $\omega\eta$ & $0$ & $0$ & $\frac{2}{3}$ & $-2$ & $\frac{2}{\sqrt{3}}$ \\
  $-$ & $\rho\pi$ & $0$ & $0$ & $-2$ & $6$ & $-2\sqrt{3}$ \\
  $-$ & $\frac{1}{\sqrt{2}}(\bar{K}^\ast K-K^\ast \bar{K})$ & $0$ & $-4\sqrt{\frac{2}{3}}$ & $\frac{2}{\sqrt{3}}$ & $-2\sqrt{3}$ & $6$ \\
  \hline
  \hline
  \end{tabular}
\end{table}

\begin{table}[htbp]
  \caption{ $D^{(2)}_{ij}$ of Eq.~(\ref{eq:nloV}) for $(S,I)=(0,1)$.} \label{coefficients 2nd term 0 1}
  \centering
  \begin{tabular}{p{20pt}p{90pt}<{\centering}p{90pt}<{\centering}p{30pt}<{\centering}p{30pt}<{\centering}p{30pt}<{\centering}p{30pt}<{\centering}p{90pt}<{\centering}}
  \hline\hline
  $G$ & & $\frac{1}{\sqrt{2}}(\bar{K}^\ast K+K^\ast \bar{K})$ & $\phi\pi$ & $\omega\pi$ & $\rho\eta$ & $\rho\pi$ & $\frac{1}{\sqrt{2}}(\bar{K}^\ast K-K^\ast \bar{K})$ \\
  \hline
  $+$ & $\frac{1}{\sqrt{2}}(\bar{K}^\ast K+K^\ast \bar{K})$ & $2$ & $0$ & $-2$ & $-\frac{2}{\sqrt{3}}$ & $0$ & $0$ \\
  $+$ & $\phi\pi$ & $0$ & $0$ & $0$ & $0$ & $0$ & $0$ \\
  $+$ & $\omega\pi$ & $-2$ & $0$ & $2$ & $\frac{2}{\sqrt{3}}$ & $0$ & $0$ \\
  $+$ & $\rho\eta$ & $-\frac{2}{\sqrt{3}}$ & $0$ & $\frac{2}{\sqrt{3}}$ & $\frac{2}{3}$ & $0$ & $0$ \\
  $-$ & $\rho\pi$ & $0$ & $0$ & $0$ & $0$ & $-4$ & $2\sqrt{2}$ \\
  $-$ & $\frac{1}{\sqrt{2}}(\bar{K}^\ast K-K^\ast \bar{K})$ & $0$ & $0$ & $0$ & $0$ & $2\sqrt{2}$ & $-2$ \\
  \hline
  \hline
  \end{tabular}
\end{table}

\begin{table}[htbp]\scriptsize
  \caption{ $D^{(3)}_{ij}$ of Eq.~(\ref{eq:nloV}) for $(S,I)=(1,\frac{1}{2})$.} \label{coefficients 6th term 1 1/2}
  \centering
  \begin{tabular}{p{40pt}p{70pt}<{\centering}p{70pt}<{\centering}p{70pt}<{\centering}p{70pt}<{\centering}p{70pt}<{\centering}}
  \hline\hline
   & $\phi K$ & $\omega K$ & $\rho K$ & $K^\ast\eta$ & $K^\ast\pi$ \\
  \hline
  $\phi K$ & $-4m_K^2$ & $0$ & $0$ & $\frac{1}{\sqrt{6}}(5m_K^2-3m_\pi^2)$ & $-\sqrt{\frac{3}{2}}(m_K^2+m_\pi^2)$ \\
  $\omega K$ & $0$ & $-2m_K^2$ & $2\sqrt{3}m_K^2$ & $\frac{1}{2\sqrt{3}}(5m_K^2-3m_\pi^2)$ & $-\frac{\sqrt{3}}{2}(m_K^2+m_\pi^2)$ \\
  $\rho K$ & $0$ & $2\sqrt{3}m_K^2$ & $-2m_K^2$ & $-\frac{1}{2}(5m_K^2-3m_\pi^2)$ & $-\frac{1}{2}(m_K^2+m_\pi^2)$ \\
  $K^\ast\eta$ & $\frac{1}{\sqrt{6}}(5m_K^2-3m_\pi^2)$ & $\frac{1}{2\sqrt{3}}(5m_K^2-3m_\pi^2)$ & $-\frac{1}{2}(5m_K^2-3m_\pi^2)$ & $-\frac{4}{3}(8m_K^2-3m_\pi^2)$ & $-2m_\pi^2$ \\
  $K^\ast\pi$ & $-\sqrt{\frac{3}{2}}(m_K^2+m_\pi^2)$ & $-\frac{\sqrt{3}}{2}(m_K^2+m_\pi^2)$ & $-\frac{1}{2}(m_K^2+m_\pi^2)$ & $-2m_\pi^2$ & $-\frac{8}{3}m_\pi^2$ \\
  \hline\hline
  \end{tabular}
\end{table}

\begin{table}[htbp]\scriptsize
  \caption{ $D^{(3)}_{ij}$ of Eq.~(\ref{eq:nloV}) for $(S,I)=(0,0)$.} \label{coefficients 6th term 0 0}
  \centering
  \begin{tabular}{p{20pt}p{70pt}<{\centering}p{70pt}<{\centering}p{70pt}<{\centering}p{70pt}<{\centering}p{70pt}<{\centering}p{70pt}<{\centering}}
  \hline\hline
  $G$ & & $\frac{1}{\sqrt{2}}(\bar{K}^\ast K+K^\ast \bar{K})$ & $\phi\eta$ & $\omega\eta$ & $\rho\pi$ & $\frac{1}{\sqrt{2}}(\bar{K}^\ast K-K^\ast \bar{K})$ \\
  \hline
  $+$ & $\frac{1}{\sqrt{2}}(\bar{K}^\ast K+K^\ast \bar{K})$ & $-6m_K^2$ & $0$ & $0$ & $0$ & $0$\\
  $-$ & $\phi\eta$ & $0$ & $\frac{32}{3}(m_\pi^2-2m_K^2)$ & $0$ & $0$ & $\sqrt{\frac{2}{3}}(5m_K^2-3m_\pi^2)$ \\
  $-$ & $\omega\eta$ & $0$ & $0$ & $-\frac{8}{3}m_\pi^2$ & $4m_\pi^2$ & $\frac{1}{\sqrt{3}}(5m_K^2-3m_\pi^2)$ \\
  $-$ & $\rho\pi$ & $0$ & $0$ & $4m_\pi^2$ & $-\frac{16}{3}m_\pi^2$ & $\sqrt{3}(m_K^2+m_\pi^2)$ \\
  $-$ & $\frac{1}{\sqrt{2}}(\bar{K}^\ast K-K^\ast \bar{K})$ & $0$ & $\sqrt{\frac{2}{3}}(5m_K^2-3m_\pi^2)$ & $\frac{1}{\sqrt{3}}(5m_K^2-3m_\pi^2)$ & $\sqrt{3}(m_K^2+m_\pi^2)$ & $-6m_K^2$ \\
  \hline
  \hline
  \end{tabular}
\end{table}

\begin{table}[htbp]\scriptsize
  \caption{ $D^{(3)}_{ij}$ of Eq.~(\ref{eq:nloV}) for $(S,I)=(0,1)$.} \label{coefficients 6th term 0 1}
  \centering
  \begin{tabular}{p{20pt}p{70pt}<{\centering}p{70pt}<{\centering}p{60pt}<{\centering}p{60pt}<{\centering}p{70pt}<{\centering}p{60pt}<{\centering}p{60pt}<{\centering}}
  \hline\hline
  $G$ & & $\frac{1}{\sqrt{2}}(\bar{K}^\ast K+K^\ast \bar{K})$ & $\phi\pi$ & $\omega\pi$ & $\rho\eta$ & $\rho\pi$ & $\frac{1}{\sqrt{2}}(\bar{K}^\ast K-K^\ast \bar{K})$ \\
  \hline
  $+$ & $\frac{1}{\sqrt{2}}(\bar{K}^\ast K+K^\ast \bar{K})$ & $-2m_K^2$ & $\sqrt{2}(m_K^2+m_\pi^2)$ & $m_K^2+m_\pi^2$ & $-\frac{1}{\sqrt{3}}(5m_K^2-3m_\pi^2)$ & $0$ & $0$ \\
  $+$ & $\phi\pi$ & $\sqrt{2}(m_K^2+m_\pi^2)$ & $0$ & $0$ & $0$ & $0$ & $0$ \\
  $+$ & $\omega\pi$ & $m_K^2+m_\pi^2$ & $0$ & $-4m_\pi^2$ & $-\frac{4}{\sqrt{3}}m_\pi^2$ & $0$ & $0$ \\
  $+$ & $\rho\eta$ & $-\frac{1}{\sqrt{3}}(5m_K^2-3m_\pi^2)$ & $0$ & $-\frac{4}{\sqrt{3}}m_\pi^2$ & $-\frac{8}{3}m_\pi^2$ & $0$ & $0$ \\
  $-$ & $\rho\pi$ & $0$ & $0$ & $0$ & $0$ & $-6m_\pi^2$ & $\sqrt{2}(m_K^2+m_\pi^2)$ \\
  $-$ & $\frac{1}{\sqrt{2}}(\bar{K}^\ast K-K^\ast \bar{K})$ & $0$ & $0$ & $0$ & $0$ & $\sqrt{2}(m_K^2+m_\pi^2)$ & $-2m_K^2$ \\
  \hline
  \hline
  \end{tabular}
\end{table}

\end{document}